\documentclass[a4paper,11pt]{article}
\pdfoutput=1 % if your are submitting a pdflatex (i.e. if you have images in pdf, png or jpg format)

\usepackage{jcappub} % for details on the use of the package, please see the JCAP-author-manual

\usepackage[T1]{fontenc}

\usepackage{color}
\usepackage{soul}% Text properties
\usepackage{graphicx}
\usepackage{dcolumn}% Align table columns on decimal point
\usepackage{bm}% bold math
\usepackage{natbib}
\usepackage{enumerate}
\usepackage{booktabs}
\usepackage{multirow}

\begin{document}

\title{A model-agnostic analysis of hybrid stars with reactive interfaces}

\author[1]{Germán Lugones,\note{Corresponding author.}}
\author[1,2,3]{Mauro Mariani,}
\author[2,3]{and Ignacio F. Ranea-Sandoval}

\affiliation[1]{Centro de Ciências Naturais e Humanas, Universidade Federal do ABC,  Avenida dos Estados 5001, CEP 09210-580, Santo André, SP, Brazil}
\affiliation[2]{Grupo de Gravitaci\'on, Astrof\'isica y Cosmolog\'ia, Facultad de Ciencias Astronómicas y Geofísicas, Universidad Nacional de La Plata, Paseo del Bosque S/N, 1900, Argentina}
\affiliation[3]{CONICET, Godoy Cruz 2290, 1425, CABA, Argentina}

% e-mail addresses: one for each author, in the same order as the authors
\emailAdd{german.lugones@ufabc.edu.br}

\abstract{We study hybrid stars considering the effects on stellar stability of the hadron-quark conversion speed at the sharp interface. The equation of state is constructed by combining a model-agnostic hadronic description with a constant speed of sound model for quark matter. We show that current LIGO/Virgo, NICER, low-density nuclear and high-density perturbative QCD constraints can be satisfied in two scenarios, with low and high transition pressures. If the conversion speed at the interface is slow, a new class of dynamically stable hybrid objects is possible and very stiff hadronic equations of state cannot be discarded.  Densities tens of times larger than the nuclear saturation density are possible at the center of these objects. We discuss possible formation mechanisms for the new class of hybrid stars and smoking guns for their observational identification.}


\maketitle

\flushbottom

%----------------------------------------------
\section{Introduction}  
%----------------------------------------------

In spite of several decades of observations and theoretical research, the nature of the deep interior of neutron stars (NSs) is still an unsolved issue.  
At present, a stringent constraint for the equation of state (EOS) comes from the observed masses of the pulsars PSR J1614-2230 \cite{Demorest:2010bx}, PSR J0348+0432 \cite{Antoniadis:2013pzd} and PSR J0740+6620  \cite{Cromartie:2019kug}, which require that an acceptable EOS must be able to support a NS of at least $2 \,M_{\odot}$.
New limits on the EOS were posed recently by the LIGO/Virgo detection of gravitational-waves (GWs) coming from the NS-NS merger event GW170817 \cite{TheLIGOScientific:2017qsa,Annala:2017llu, Most:2018hfd, Raithel:2018ncd, Capano:2019eae}. Assuming that both NSs are described by the same EOS and have spins within the range observed in Galactic binary NSs,  the dimensionless tidal deformability $\Lambda_{1.4}$ of a $1.4 \,M_{\odot}$ NS was found to be in the range $70-580$ at the $90 \%$ level  \cite{Abbott:2018exr}.
Also, the fact that the postmerger remnant of GW170817 did not suffer a prompt collapse was used to constrain the maximum gravitational mass of a non-rotating NS to be $2.17^{+0.17}_{-0.15}\,M_\odot$ \cite{Rezzolla_2018}.
The mass and radius of the merging objects in the GW190425 event were also inferred,  but the possibility that one or both components are black holes cannot be ruled out  \cite{Abbot:2020goo}.

Additionally, the Neutron Star Interior Composition Explorer (NICER) has measured the mass and radius of the millisecond-pulsars PSR J0030+0451 \cite{Riley:2019yda,Miller:2019cac} and PSR J0740+6620 \cite{Riley:2021pdl,Miller:2021qha} with great precision.
Before the radius measurement of PSR J0740+6620, a comparison of observations with a large variety of EOSs led to the idea that both extremely stiff and soft matter would be ruled out \cite{Abbott:2018exr,Capano:2019eae}. 
However, the latest joint NICER and XMM-Newton observation showed that the  radius of PSR J0740+6620 ($M \approx 2\,M_{\odot}$)  is very similar to that of PSR J0030+0451  ($M\approx 1.4\,M_{\odot}$), even though they have very different masses \cite{Riley:2021pdl,Miller:2021qha}. 
These results favor stiff EOSs \cite{Raaijmakers:2021uju} and create some tension with the masses and radii inferred for the objects in GW170817.

For decades, microscopic theories of matter have tried to reveal the EOS of NS interiors. At present, the EOS is well founded on nuclear theory and  experiments for densities $\lesssim 1 \, n_0$ (being $n_0=0.16~$fm$^{-3}$, the nuclear saturation density). Beyond $\sim 40\,n_0$, perturbative QCD (pQCD) can describe deconfined quark matter accurately \cite{Kurkela:2009gj, Gorda:2018gpy},  but such large densities are not usually expected in NSs.  Between these limits, a robust approach is to use a set of model-agnostic EOSs interpolating both regimes without violating causality, and requiring that the resulting NS configurations fulfil astrophysical constraints. However, it is still unclear at which density would a hadron-quark transition occur and whether hybrid stars (HSs) containing quark cores would exist in Nature.

Concerning HSs, it is under debate whether quarks and hadrons are separated by a sharp discontinuity or by a mixed phase where they coexist along a wide density region forming globally charge-neutral geometrical structures. Mixed phases are energetically preferred if the quark matter surface tension $\sigma$ is smaller than a critical value $\sigma_\mathrm{crit}$ of the order of tens of $\mathrm{MeV} / \mathrm{fm}^{2}$; but if $\sigma > \sigma_\mathrm{crit}$, the mixed phase is unstable and a sharp interface is favored \cite{Wu:2018zoe,Maslov:2018ghi}. 
Unfortunately, theoretical values of $\sigma$ span a wide range depending on the EOS and on the calculation method:  some authors  obtain  $\sigma < \sigma_\mathrm{crit}$ \cite{Garcia:2013eaa,Lugones:2016ytl,Lugones:2018qgu,Fraga:2018cvr,Gao:2016hks} but  very large $\sigma$ favoring a sharp interface is obtained using the multiple reflection expansion method with the Nambu-Jona-Lasinio EOS  \cite{Lugones:2013ema} and the MIT bag EOS with vector interactions \cite{Lugones:2021tee}.
In this work, we adopt as a working hypothesis that the  interface is sharp. 

On the other hand, it has been shown that the conversion speed between quarks and hadrons at a sharp interface in a HS is deeply related to the dynamic stability of the object \cite*{VasquezFlores:2012vf, Pereira:2017rmp}. In fact, if conversions have a sufficiently long timescale compared to the typical oscillation timescale of the star ($\sim 1 \, \mathrm{ms}$), a new branch of stable hybrid stars is possible. In this work, we will explore systematically this possibility and will analyse several physical and astrophysical consequences of their hypothetical existence.

The paper is organized as follows. In Section \ref{sec:2}, we summarize the role of interface reactions on the dynamic stability of hybrid stars emphasizing that a new class of compact stars is possible if the reaction timescale is slower that the typical oscillation frequency.  In Section \ref{sec:3}, we discuss the microphysics of the hadron-quark conversion in degenerate matter and present arguments supporting the idea that it may be slow.  In Section \ref{sec:4}, we describe the EOS that will be used in our calculations. In Section \ref{sec:5}, we present our results for the mass-radius relationship and the tidal deformability of the new class of hybrid objects, showing that they are in agreement with current observations. In Section \ref{sec:6}, we explore some physical and astrophysical consequences of our results and give some prospects for the detection of the new class of objects.

%-----------------------------------------------------------------
\section{The role of interface reactions on stellar stability}
\label{sec:2}
%-----------------------------------------------------------------

As shown by Chandrasekhar \cite{Chandrasekhar:1964zz},  stellar stability can be assessed by inspecting the response of equilibrium configurations to small radial perturbations. In a dynamically unstable star, small perturbations grow without limit, leading to the collapse or disruption of the object. 
In a stable star, fluid elements along the stellar interior oscillate around their equilibrium positions, compressing and expanding periodically. The formalism of small radial perturbations of spherically symmetric stars shows that stellar configurations are stable if the frequency $\omega_{0}$ of the fundamental mode verifies $\omega_{0}^2 \geq 0$ and unstable if $\omega_{0}^2 < 0$. 

However, in the case of HSs, Chandrasekhar's analysis is not straightforward because radial perturbations may induce phase conversions in the neighborhood of the interface.  
Let us first assume that phase conversions are \emph{slow}, i.e. that the conversion timescale $\tau_{\mathrm{conv}}$ turns out to be much larger than the oscillation period $\tau_{\mathrm{osc}} = 2 \pi \omega_0^{-1}$ of perturbed fluid elements (notice that for typical NSs, $\tau_{\mathrm{osc}} \sim 1 \, \mathrm{ms}$).  When matter close to the quark-hadron interface is disturbed and radially displaced from its equilibrium position, it will maintain its  composition even if its pressure fluctuates above and below the transition pressure, $p_t$. Thus, the interface oscillates around its unperturbed position and fluid elements on either side cannot traverse it.
Since the interface oscillates with the same period as the disturbances,  its movement can be encoded in junction conditions for the relative radial displacement, $\xi$, and the Lagrangian perturbation of the pressure, $\Delta p$, which must be continuous across the phase splitting surface \cite*{Pereira:2017rmp}:
\begin{equation}
[\xi ]^+_{-}\equiv \xi^+-\xi^- =0 \,,\qquad [\Delta p]^+_- \equiv \Delta p^+ - \Delta p^- =0 \,,
\label{xislow}
\end{equation}
where $^+$ and $^-$ indicate the function values on each side of the interface. 
Otherwise, if conversions are \emph{rapid} ($\tau_{\mathrm{conv}} \ll \tau_{\mathrm{osc}}$), the  interface stays at rest with respect to the stellar center and matter moves across it changing {\emph{instantaneously}} its composition. In this case  $[\xi-  \Delta p / (r_t p_0^\prime) ]$ and $\Delta p$ are continuous across the interface,
\begin{equation}
[\xi]^+_-=\Delta p\left [\frac{1}{r_t p'_0}\right]^+_- \, ,  \qquad \qquad [\Delta p]^+_- =0  ,
\label{xirapid}
\end{equation}
being $r_t$ its radial position and  $p_0^\prime$ the pressure's radial derivative in the unperturbed configuration \cite*{Pereira:2017rmp}\footnote{A non-relativistic version of Eqs. \eqref{xislow} and  \eqref{xirapid} was first derived in Ref. \cite{Haensel1989}. Similar relativistic conditions were obtained in Ref. \cite{Karlovini:2003xi} in the context of elastic stars. However, no connection between Eq. \eqref{xislow} (slow conversions) and extended stellar stability was established in these works. }.

\begin{figure}
\centering
\includegraphics[width=0.66\linewidth,angle=0]{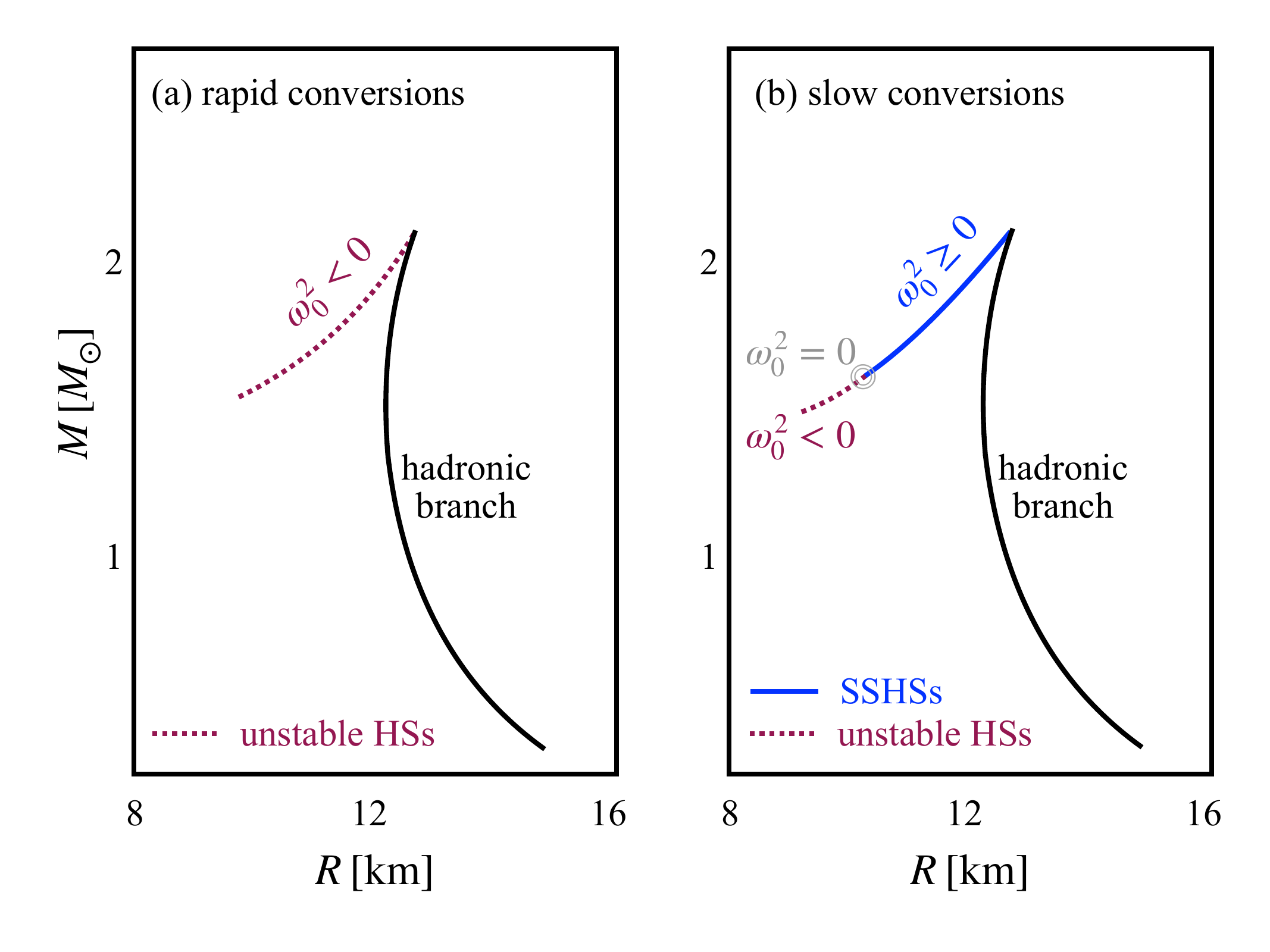}
\caption{Schematic representation of the stability of hybrid stars with sharp interfaces in a scenario where the transition density is high. In both panels the EOSs for hadronic and quark matter are the same, as well as the transition density. The only difference is the junction condition used at the interface to describe (a) rapid and (b) slow conversions when the star is radially perturbed. SSHSs are dynamically stable ($\omega_{0}^2 \geq 0$) even if  $\partial M/\partial \epsilon_c < 0$. }
\label{fig:SSHSs-1}
\end{figure}

The existence of junction conditions embodying the conversion speed has strong consequences for the dynamic stability of hybrid stars. In particular, it can be shown that in the case of slow conversions, the maximum mass star is no longer the one that separates stable configurations from unstable ones, as we explain below.
For cold catalyzed stars, it is known that changes of stability ($\omega_{0}=0$)  occur at maxima or minima in the $M-\epsilon_c$ diagram, being $\epsilon_c$ the energy density at the NS center. Thus, the following static stability criterion holds \cite{Harrison1965}:
\begin{eqnarray}
    \frac{\partial M}{\partial \epsilon_c}<0 \quad &\Rightarrow& \quad \omega_{0}^2 < 0 \quad  \text{(unstable star)},      \label{dMdE_c_1}  \\
    \frac{\partial M}{\partial \epsilon_c}>0 \quad  &\Leftarrow&  \quad  \omega_{0}^2 \geq 0 \quad  \text{(stable star)} .  
    \label{dMdE_c_2}
\end{eqnarray}
However, when chemical reactions are possible in some part of the star, matter is not necessarily in thermodynamic equilibrium everywhere and Eqs.~\eqref{dMdE_c_1} and \eqref{dMdE_c_2} may fail \footnote{Some examples of the failure of Eqs. \eqref{dMdE_c_1} and \eqref{dMdE_c_2} are already known in the literature in different contexts; for example, in purely hadronic stars with frozen oscillations \cite{Gourgoulhon}, in electrically charged strange quark stars \cite{Arbanil:2015uoa}, and in compact objects with dark matter  (see e.g. Fig.~2 of \cite{Kain:2020zjs}).}. 
More specifically, if interface conversions are slow, matter in the neighborhood of the phase splitting surface is non-catalyzed (there is not enough time to attain equilibrium) and the presumptions that lead to Eqs.~\eqref{dMdE_c_1} and \eqref{dMdE_c_2} are not fulfilled.
In fact, numerical calculations have shown that in the case of HSs with sharp density discontinuities  and \textit{slow interface conversions}, $\omega_{0}$ can be a real number (indicating stability) even if $\partial M/\partial \epsilon_c < 0$.
Stable stellar configurations with $\partial M/\partial \epsilon_c < 0$ were found in Ref. \cite{VasquezFlores:2012vf} and their connection with slow interface conversions was first established in Ref. \cite{Pereira:2017rmp}. Further investigations \cite{Mariani:2019vve, Tonetto:2020bie,  Parisi:2020qfs,DiClemente2020,Pereira:2020cmv, Rodriguez:2020fhf, mariani2022MNRAS, Ranea:2022bou} explored different aspects of these new configurations for specific EOSs but a systematic  analysis for a more general family of EOSs was still lacking and will be addressed in the present work.

\begin{figure}
\centering
\includegraphics[width=0.66\linewidth,angle=0]{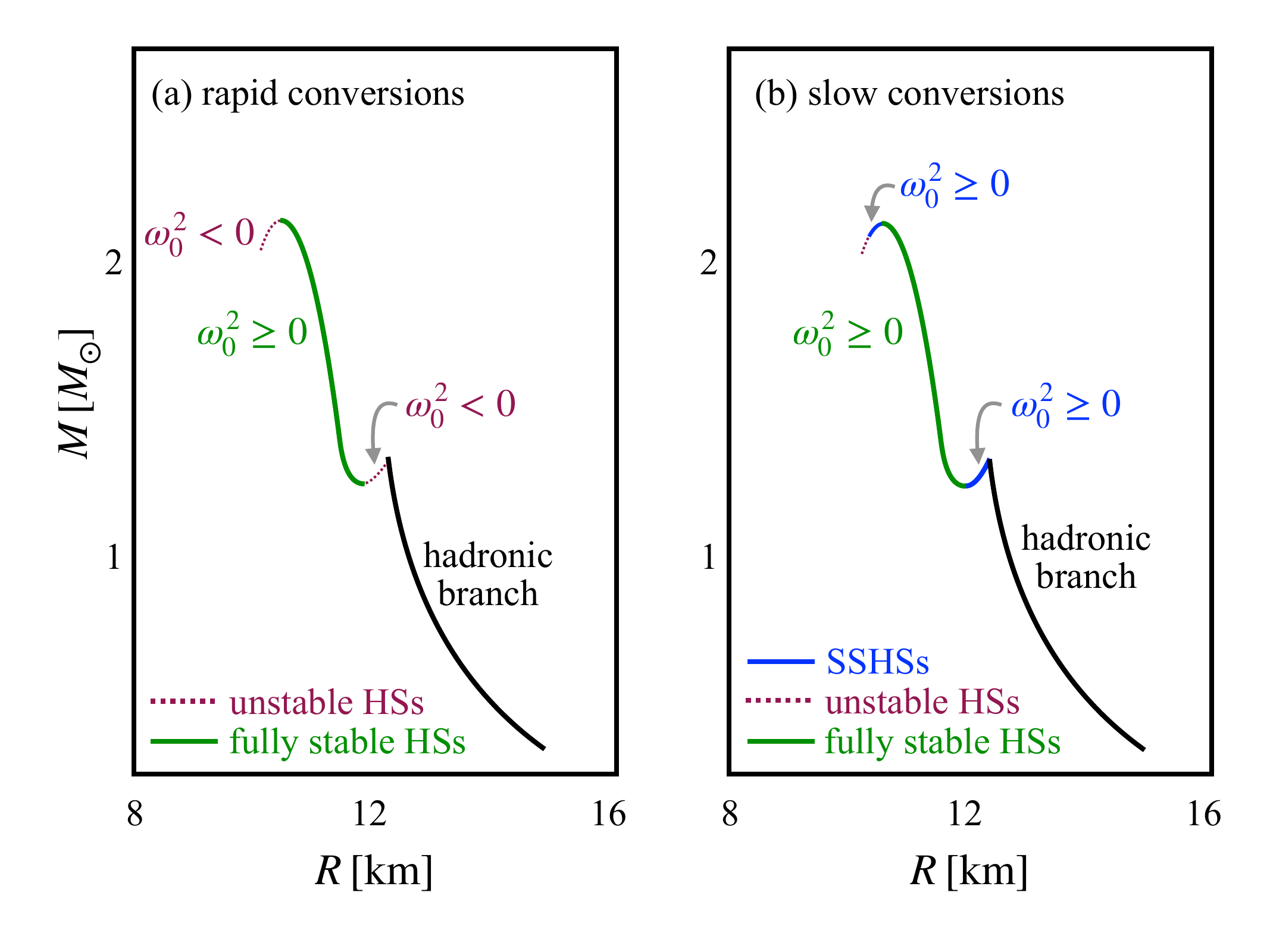}
\caption{Same as in the previous figure but in a scenario where the transition density is low. The green  branch (fully stable HSs)  is stable for any conversion speed (rapid or slow). As in the previous figure, the blue branches (SSHSs) are stable only for slow conversions. }
\label{fig:SSHSs-2}
\end{figure}

HSs that are stable even if $\partial M/\partial \epsilon_c < 0$ will be called  \emph{slow-stable} (SS) configurations or SS hybrid stars (SSHSs). A schematic representation of SSHSs in the mass-radius diagram is shown in Figs. \ref{fig:SSHSs-1} and  \ref{fig:SSHSs-2}. When the transition density is high enough, they usually arise to the left of the maximum mass object, i.e. their central densities are larger than the central density of the star with $M_\mathrm{max}$ (see Fig. \ref{fig:SSHSs-1}b). If the hadron-quark transition occurs at a low enough density, configurations like the one shown in Fig. \ref{fig:SSHSs-2}(b) are possible. Although other possible forms of the HS mass-radius relation may arise when a single first-order phase transition is taken into account \cite{Alford:2013aca},  in this work we will focus mainly on the case represented in Fig. \ref{fig:SSHSs-1}(b), due to its important physical and astrophysical consequences.

Strictly speaking, the SSHS branch does not extend exactly up to the star with $\omega_{0}^2=0$ ($\tau_{\mathrm{osc}}=\infty$), but up to an object with a smaller central density that has $\omega_{0}^2 \sim  2 \pi \tau^{-1}_{\mathrm{conv}}$   (i.e. $\tau_{\mathrm{osc}} \sim \tau_{\mathrm{conv}}$). 
However, in practice $\omega_0$ is of the order of a few kilohertz for almost all HSs in the SS branch, and tends steeply to zero very close to the mass of the zero-frequency object. This behavior can be verified in Fig.~$11$ of Ref. \cite{Pereira:2017rmp}, where it is seen that $\omega_0$ has a significantly large value (some fraction of $1$~kHz) even for HSs that are extremely close to the zero-frequency object. 
Therefore, if  $\tau_{\mathrm{conv}}$ is larger than some milliseconds,  the difference between locating the stable-unstable boundary at the zero-frequency point ($\tau_{\mathrm{osc}} = \infty$) or at the point with  $\tau_{\mathrm{osc}} \approx \tau_{\mathrm{conv}}$ is negligible. 
If $\tau_{\mathrm{conv}} \sim 1 \mathrm{ms}$, the stable-unstable border would be shifted to the right in Fig. \ref{fig:SSHSs-1}(b), resulting in a shorter SSHS branch. As  $\tau_{\mathrm{conv}}$  decreases below $\sim 1 \, \mathrm{ms}$, the border gets closer and closer to the maximum of the $M-R$ curve. In the limiting case of   $\tau_{\mathrm{conv}} = 0$ (infinitely rapid conversions)  the border coincides with the maximum, Eqs.~\eqref{dMdE_c_1}-\eqref{dMdE_c_2} become valid again, and the SSHS branch would not exist. This is easy to understand because for  $\tau_{\mathrm{conv}} = 0$ perturbed matter is in thermodynamic equilibrium at all times, i.e. the assumption of catalyzed matter that led to Eqs.~\eqref{dMdE_c_1}-\eqref{dMdE_c_2} \cite{Harrison1965} is always fulfilled.

\begin{figure}
\centering
\includegraphics[width=0.66\linewidth,angle=0]{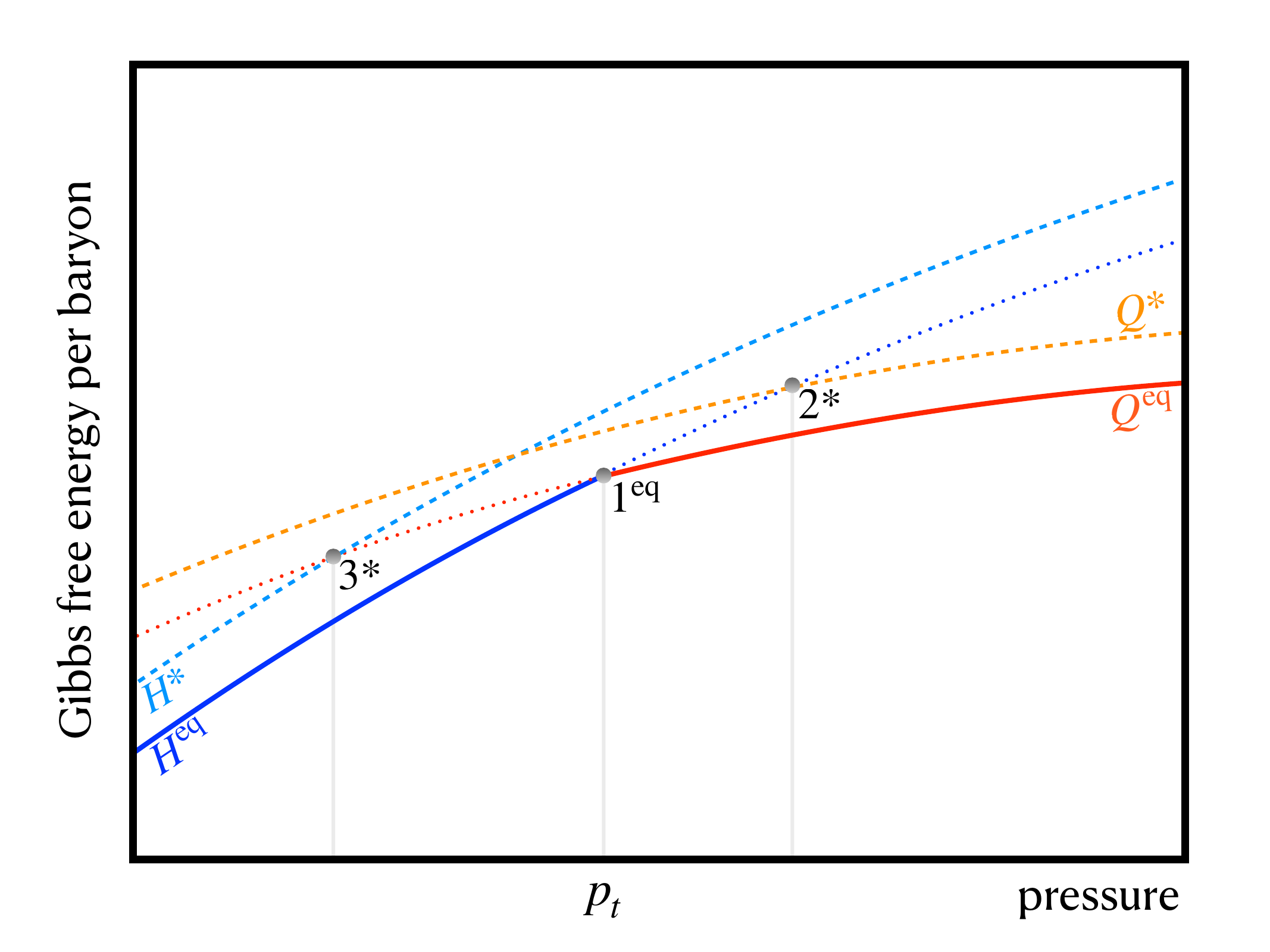}
\caption{Schematic representation of the Gibbs free energy per baryon as a function of pressure for hadron and quark matter in chemical equilibrium ($H^{\mathrm{eq}}$ and $Q^{\mathrm{eq}}$) and for the non-catalyzed transition states ($H^*$ and $Q^*$). The quark phase $Q^{*}$ is out of chemical equilibrium and is defined as the one that has the same flavor composition than $H^{\mathrm{eq}}$ at the same pressure (see Refs. \cite{Olesen:1993ek, Iida:1998pi,  Lugones:1997gg, Bombaci:2004mt, Bombaci:2007eoc, Lugones:2015bya} for more details).  Analogously, $H^{*}$ has the same flavor composition than $Q^{\mathrm{eq}}$ at the same $p$. The HS interface is located at the point $1^{\mathrm{eq}}$ with a pressure $p_t$.  A hadronic fluid element with a pressure slightly below $p_t$ will deconfine if a perturbation is able to take it to the point $2^*$ at which $H^{\mathrm{eq}}$ and $Q^*$ have the same free energy. Similarly, a quark fluid element with a pressure slightly above $p_t$ will hadronize only if it reaches the point $3^*$ after a perturbation. Quantum and thermal nucleation timescales for these transitions are  extremely long at low temperatures (see text).   }
\label{fig:meta}
\end{figure}

%-----------------------------------------------------------------
\section{The speed of interface conversions}
\label{sec:3}
%-----------------------------------------------------------------

According to the previous discussion, interface conversions in a HS are slow if $\tau_{\mathrm{conv}} \gg 1 \,\mathrm{ms}$, and  fast if $\tau_{\mathrm{conv}} \ll 1 \, \mathrm{ms}$. 
Unfortunately,  the actual conversion timescale in the hypothetic sharp interface of a HS is uncertain.  Nonetheless, the transition timescale is not expected to be simply the result of particles that interact independently, because phase transitions are highly collective and nonlinear phenomena. In other words, although typical interaction timescales are $\sim 10^{-23} \,\mathrm{s}$ for the strong force and $\sim 10^{-8}\, \mathrm{s}$ for the weak one, the conversion timescale in a HS is not necessarily that fast. In principle, one expects that the conversion would be driven by nucleation or spinodal decomposition, since these are the two major ways by which first-order phase transitions proceed in a wide variety of systems (see e.g. Ref. \cite{Slezov2009}).  However, other mechanisms based on strangeness diffusion have been proposed \cite{Olinto:1986je, Alford:2014jha}.

In the case of nucleation, there are quantum and thermal nucleation models describing the case where there is no pre-existing quark matter, and quark drops must nucleate somewhere in the high-density region of the NS (see Ref. \cite{Bombaci:2016xuj} and references therein). The same analysis can be applied to the present case where a pre-existing quark-hadron interface moves in response to a density fluctuation. The mechanism can be understood by analyzing the Gibbs free energy per baryon $G/n_B$  of the different phases involved in the transition (see Fig. \ref{fig:meta}). Catalyzed hadronic matter in the neighborhood of the interface (point $1^\mathrm{eq}$) will deconfine only if it reaches the point $2^*$, while quark matter close to the point $1^\mathrm{eq}$ will convert into hadrons only if it reaches the point $3^*$. A direct conversion $H^{\mathrm{eq}} \leftrightarrow  Q^{\mathrm{eq}}$  is strongly suppressed because both phases have in general a very different flavor composition and a high-order weak interaction process would be needed \cite{Lugones:2015bya, Bombaci:2007eoc,Lugones:1997gg,Iida:1998pi,Olesen:1993ek}. Quantum and thermal nucleation timescales for the $1^\mathrm{eq} \rightarrow 2^*$ transition are typically orders of magnitude larger than the age of the Universe for temperatures below a few MeV \cite{Bombaci:2016xuj}. We are not aware of calculations of the $1^\mathrm{eq} \rightarrow 3^*$ transition, but similar timescales can be reasonably expected.

Finally, a conversion mechanism based on strangeness diffusion has  been considered \cite{Alford:2014jha} and seems powerful enough to saturate $r$-modes at very low amplitude, of order $10^{-10}$. A similar saturation can be expected for radial oscillations. 

The above discussion shows that although the conversion mechanism is not fully understood, slow conversions represent a feasible scenario. Moreover, since in the limit of $\tau_{\mathrm{conv}} \ll 1 \, \mathrm{ms}$ and $\tau_{\mathrm{conv}} \gg 1 \, \mathrm{ms}$ the stellar response is  independent of the kinetic details of the conversion (they can be encoded in junction conditions), we can obtain robust conclusions about stellar stability in spite of the microphysical uncertainties of the phase transformation, whose understanding is not the objective of this work.

%-----------------------------
\section{Hybrid EOS Model}
\label{sec:4}
%-----------------------------

\begin{table}[tb]
\footnotesize
\centering
\begin{tabular}{ccccccccc}
\toprule
\# & \multicolumn{4}{c}{Hadronic EOS} & &   \multicolumn{3}{c}{Quark EOS} \\ 
\cmidrule(lr){2-5}\cmidrule(lr){7-9}
 & $\log_{10}\rho_1 (\mathrm{g/cm^3})$ & $\log_{10}\rho_2 (\mathrm{g/cm^3})$  & $\Gamma_2$  & $\Gamma_3$ & & $p_{t} \left[{\rm {MeV}/{fm^{3}}}\right]$   &    $\Delta \epsilon \left[{\rm {MeV}/{fm^{3}}}\right]$  &  $c^2_\textrm{s}$  \\
\toprule
1 & 14.43 & 14.58 & 5.9 & 2.0 & & 150 & 1800 & 0.33 \\ \midrule
2 & 14.43 & 14.58 & 6.2 & 2.3 & & 150 & 3000 & 0.33 \\ \midrule
3 & 14.45 & 14.58 & 6.5 & 2.6 & & 150 & 2000 & 0.30 \\ \midrule
4 & 14.45 & 14.68 & 6.5 & 2.6 & & 70 & 1000 & 0.33 \\\midrule
5 & 14.45 & 14.58 & 8.5 & 2.0 & & 20 & 100 & 0.50 \\\midrule
6 & \multirow{4}{*}{14.45} & \multirow{4}{*}{14.58} & \multirow{4}{*}{10.0} & \multirow{4}{*}{2.0} & & 10 & 100 & 0.50 \\
\cmidrule{1-1} \cmidrule{7-9}
7 & & & & & & 60 & 600 & 0.20 \\
\cmidrule{1-1} \cmidrule{7-9}
8 & & & & & & 60 & 900 & 0.33 \\
\bottomrule
\end{tabular}
\caption{Parameters of the selected hybrid EOSs. In all cases, we adopted  $\log_{10} \rho_0 (\mathrm{g/cm^3}) = 13.980$, $\log_{10}K_1 = -27.22$ and $\Gamma_1= 2.764$ for the first piece of the hadronic core EOS to match the upper limit predicted by cEFT EOSs (see Fig.~\ref{fig:equation-of-state}). The values of $\rho_1$ and $\rho_2$ were changed slightly from the ones originally given in Ref. \cite{OBoyle-etal-2020}.}  
\label{tabla:param_selec}
\end{table}

For the hadronic crust we use a generalized piecewise polytropic (GPP) fit \cite{OBoyle-etal-2020}  to the SLy(4) EOS found in Ref. \cite{Douchin:2001sv}, which accurately reproduces the EOS and the adiabatic index. 
For the hadronic part of the core, we construct several different model-agnostic GPP EOSs using the prescription of Ref. \cite{OBoyle-etal-2020}.
This part of the EOS is divided into three segments by three specific densities. The first boundary, located at $\rho_0$, marks the transition between the crust and the lowest-density segment of the core. The dividing densities $\rho_1$ and $\rho_2$ separate the remaining segments and their values are listed in Table \ref{tabla:param_selec}. These densities were obtained in Ref. \cite{OBoyle-etal-2020} by minimizing an error norm based on integral astrophysical observables. Between the dividing densities, the  rest-mass density $\rho$, energy density $\epsilon$, and speed of sound $c_s$ are expressed by the following functions of the pressure $p$ \footnote{As in Refs. \cite{Hebeler:2013nza, OBoyle-etal-2020}, we adopt the convention of incorporating the speed of light $c$ into the definition of energy density and pressure. Consequently, the EOS parameters are given in units such that $\rho$,  $\epsilon$, and  $p$ are expressed in  $\mathrm{g/cm}^{3}$.}:
\begin{eqnarray}
\rho &=& \left(\frac{p - \Lambda_i}{K_i}\right) ^ {\frac{1}{\Gamma_i}}   ,   \\
\epsilon &=& \frac{K_i \rho ^ \Gamma_i}{\Gamma_i - 1} + (1 + a_i) \rho - \Lambda_i   ,  \\
c_s &= & \left[\frac{1}{\Gamma_i - 1} + \frac{1 + a_i}{K_i \Gamma_i \rho ^ {\Gamma_i - 1}}\right] ^ {-\frac{1}{2}}   .
\end{eqnarray}
The set of parameters $K_i$, $\Gamma_i$, $a_i$, and $\Lambda_i$ characterize the EOS within each interval $[\rho_{i-1}, \rho_i]$. These parameters are not independent, as continuity and differentiability of the energy density $\epsilon(\rho)$ and pressure $p(\rho)$ are enforced across the dividing densities. In Ref. \cite{OBoyle-etal-2020}, the parameters $\rho_0$, $K_1$, $\Gamma_{1}$, $\Gamma_{2}$, and $\Gamma_{3}$ were treated as independent, while the other parameters were determined from the junction conditions described previously (see Ref. \cite{OBoyle-etal-2020} for more details). Here, we adopted $\log_{10} \rho_0 (\mathrm{g/cm^3}) = 13.980$, $\log_{10}K_1 = -27.22$ and $\Gamma_1= 2.764$ for the first piece of the hadronic core EOS to match at $1.1 \,n_0$ to the upper limit of a calculated EOS range based on chiral effective field theory (cEFT) interactions including theoretical uncertainties (see Ref. \cite{Hebeler:2013nza} for details).
The other EOS parameters were chosen arbitrarily to obtain ultrastiff EOSs that fulfill causality and are consistent with $2\,M_{\odot}$ pulsars. 
Yet, these stiff EOSs are intentionally built to produce \emph{hadronic} $M$-$R$ curves that do not satisfy the constraints from GW170817  \cite{Annala:2017llu,Annala:2020efq}.
Finally, note that since both $p(\rho)$ and $\epsilon(\rho)$ are differentiable functions, the hadronic sound speed $c_{s}$ is continuous at these boundaries.

%%%%%%%%%%%%%%%% FIGURE 1
\begin{figure}
\centering
\includegraphics[width=0.6\linewidth,angle=0]{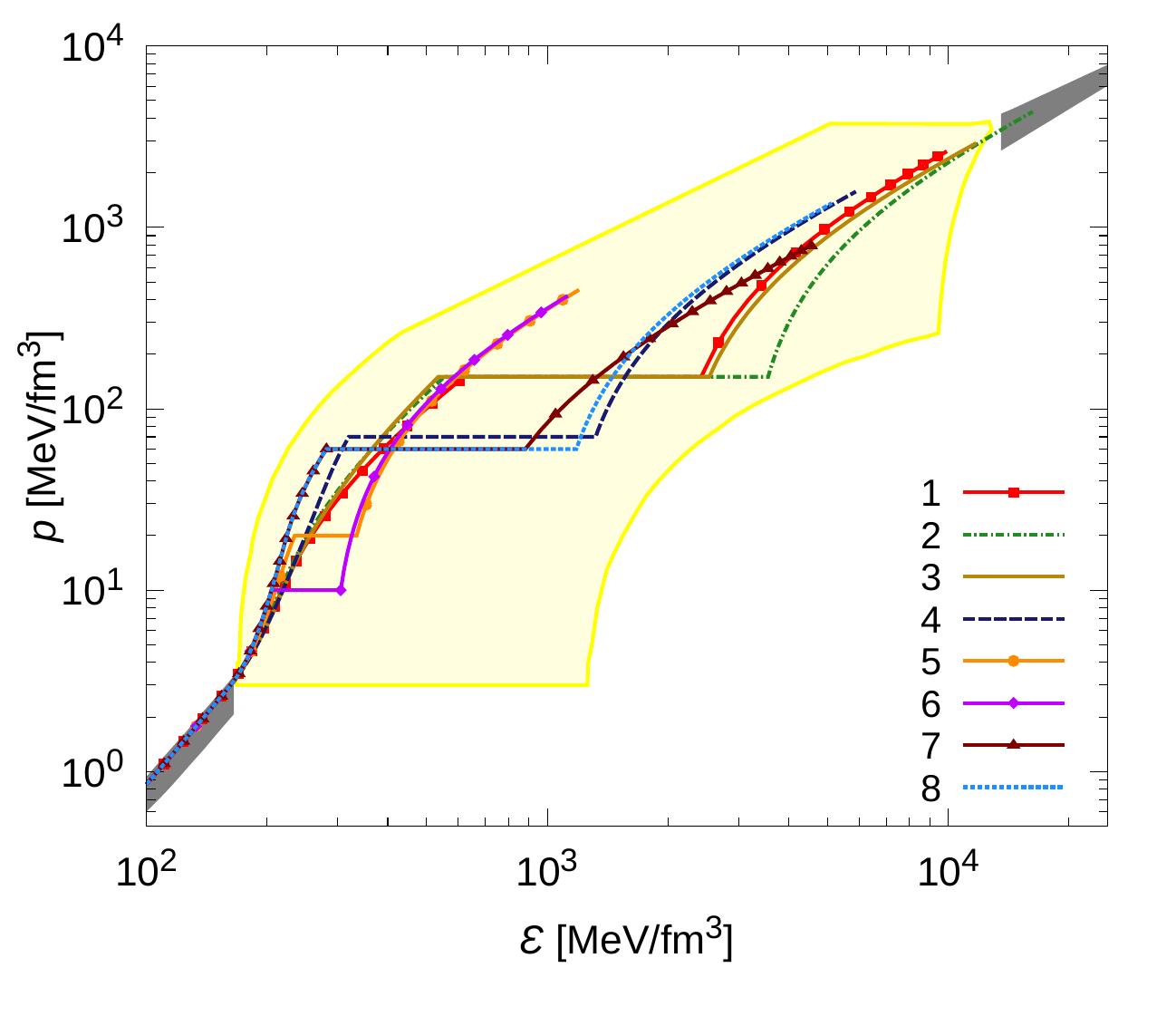}
\caption{Selected hybrid EOSs constructed with \emph{stiff} hadronic EOSs ruled out by GW170817 original data analysis but fulfilling NICER and $2~M_{\odot}$ pulsar observations.  EOSs are constrained by calculations using cEFT Hamiltonians up to $1.1\, n_0$ and perturbative QCD results for $n \gtrsim 40\, n_0$ (grey bands, see \cite{Hebeler:2013nza, Kurkela:2009gj, Gorda:2018gpy}). The yellow region shows the recent constraint obtained in Ref. \cite{Komoltsev:2022hpq}.  Curves are displayed up to the largest central density in each HS model. More details of these hybrid EOSs are shown in Table~\ref{tabla:param_selec}. } 
\label{fig:equation-of-state}
\end{figure}

We assume that at a given pressure $p_{t}$ a first-order phase transition between hadronic and quark matter occurs, with both phases separated by a sharp interface. To keep our analysis as general as possible, quark matter is described by the constant speed of sound model \cite{Alford:2013aca}, which is parametrized in terms of $p_{t}$, the energy density jump $\Delta \epsilon$ between both phases, and  $c_\mathrm{s}$ (assumed to be constant). 
The phase transition is described by the Maxwell construction, which guarantees that the pressure and the Gibbs free energy per baryon are the same at both sides of the interface. 
In this way, we constructed over 3000 hybrid EOSs using parameters in the following ranges: 
\begin{eqnarray}
    & 10~ \mathrm{MeV/fm^3} \le p_{t} \le 300~\mathrm{MeV/fm^3} , \\
    & 100~\mathrm{MeV/fm^3} \le \Delta \epsilon \le 3000~\mathrm{MeV/fm^3} , \\
    & 0.2 \le c_\mathrm{s}^2 \le 1  ,
\end{eqnarray}
checking consistency with pQCD calculations for \mbox{$n \gtrsim 40\, n_0$} \cite{Kurkela:2009gj, Gorda:2018gpy} and  picking only EOSs whose stellar models verify the maximum mass constraints  \mbox{$2.01^{+0.04}_{-0.04}  < M_{\max}/M_\odot  < 2.16^{+0.17}_{-0.15}$} \cite{Rezzolla_2018}, $M_\textrm{max} \leq 2.3~M_\odot$ \cite{Shibata:2019ctb}, and \mbox{$2.15^{+0.18}_{-0.17}  < M_\textrm{max} / M_\odot < 2.24^{+0.45}_{-0.44}$} \cite{Khadkikar:2021yrj}.
From these results, we selected eight representative hybrid EOS (see Fig.~\ref{fig:equation-of-state} and  Table~\ref{tabla:param_selec}) which satisfy all recent astrophysical constraints and exemplify qualitatively the broad range of results we obtained.

%%%%%%%%%%%%%%%% FIGURE 2
\begin{figure*}
\centering
\includegraphics[width=0.6\linewidth,angle=0]{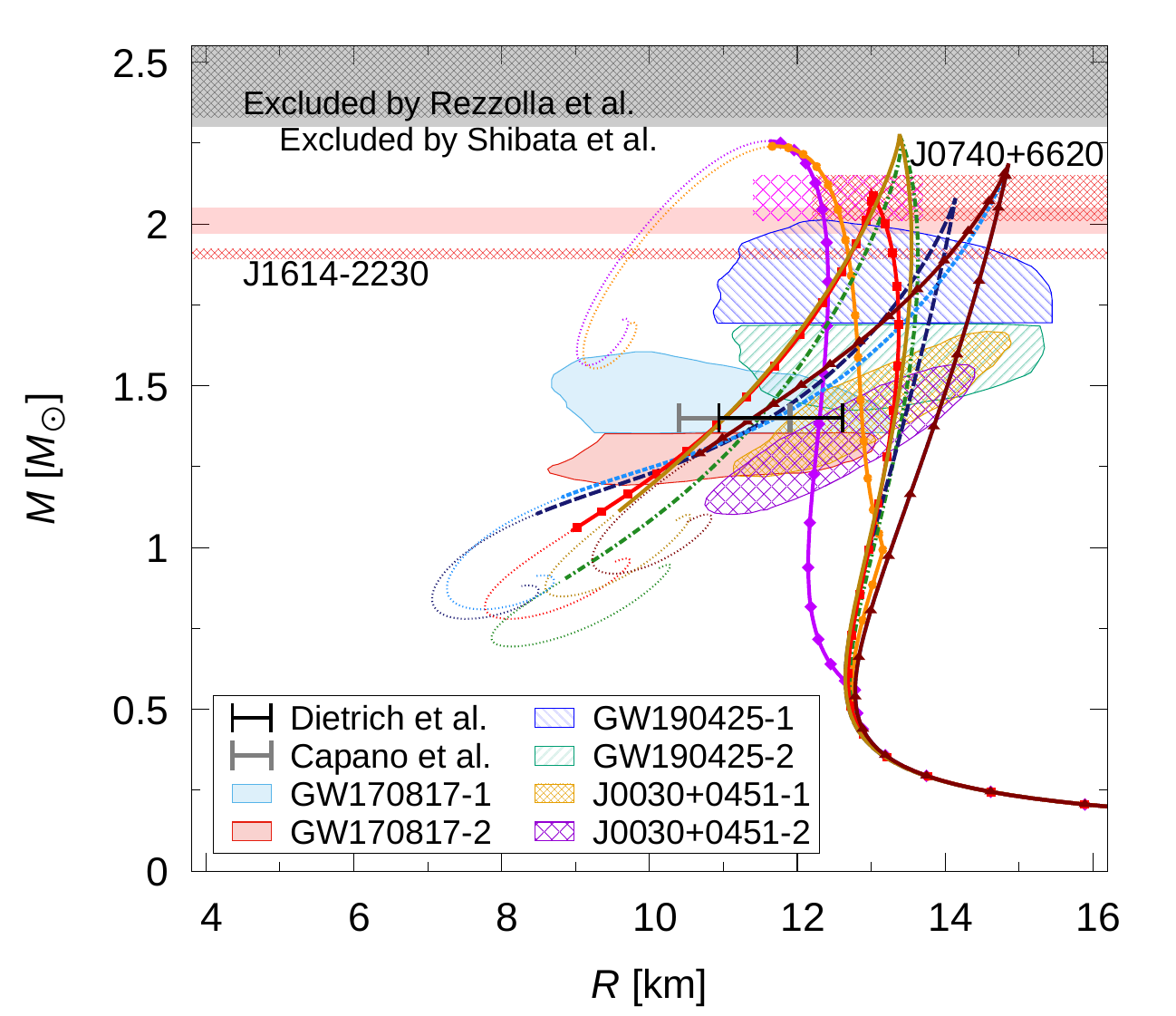}
\caption{Mass-radius relationships for the EOSs of Fig.~\ref{fig:equation-of-state} (same color-coding and symbols are used). We also show astrophysical constraints from the \mbox{$\sim 2~M_\odot$} pulsars, GW170817 \cite{Raithel:2018ncd, Annala:2017llu}, GW190425 \cite{Abbot:2020goo}, and NICER \cite{Riley:2019yda,Miller:2019cac} observations. In gray and black, we show radius constraints from the analysis of \cite{Dietrich:2020mco} and \cite{Capano:2019eae}, respectively. In hatched gray, the region excluded due to the constraint $M_\textrm{max} \leq 2.16^{+0.17}_{-0.15}~M_\odot$ of Ref. \cite{Rezzolla_2018} and in solid light gray the one excluded by Ref. \cite{Shibata:2019ctb}, $M_\textrm{max} \leq 2.3~M_\odot$. The condition $M_\textrm{max} \leq 2.24^{+0.45}_{-0.44}~M_\odot$ of Ref. \cite{Khadkikar:2021yrj} is not  displayed  in the figure for clarity, since it lies far outside the $y$-axis range. When slow conversions are assumed, all points of the thick curves represent stable stars and if conversions are rapid some hybrid segments are unstable (see text). Thin lines represent unstable stars in both scenarios.}
\label{fig:MR}
\end{figure*}

%--------------------------------
\section{Results}
\label{sec:5}
%--------------------------------

%%%%%%%%%%%%%%%% FIGURE 3
\begin{figure}
\centering
\includegraphics[width=0.6\linewidth,angle=0]{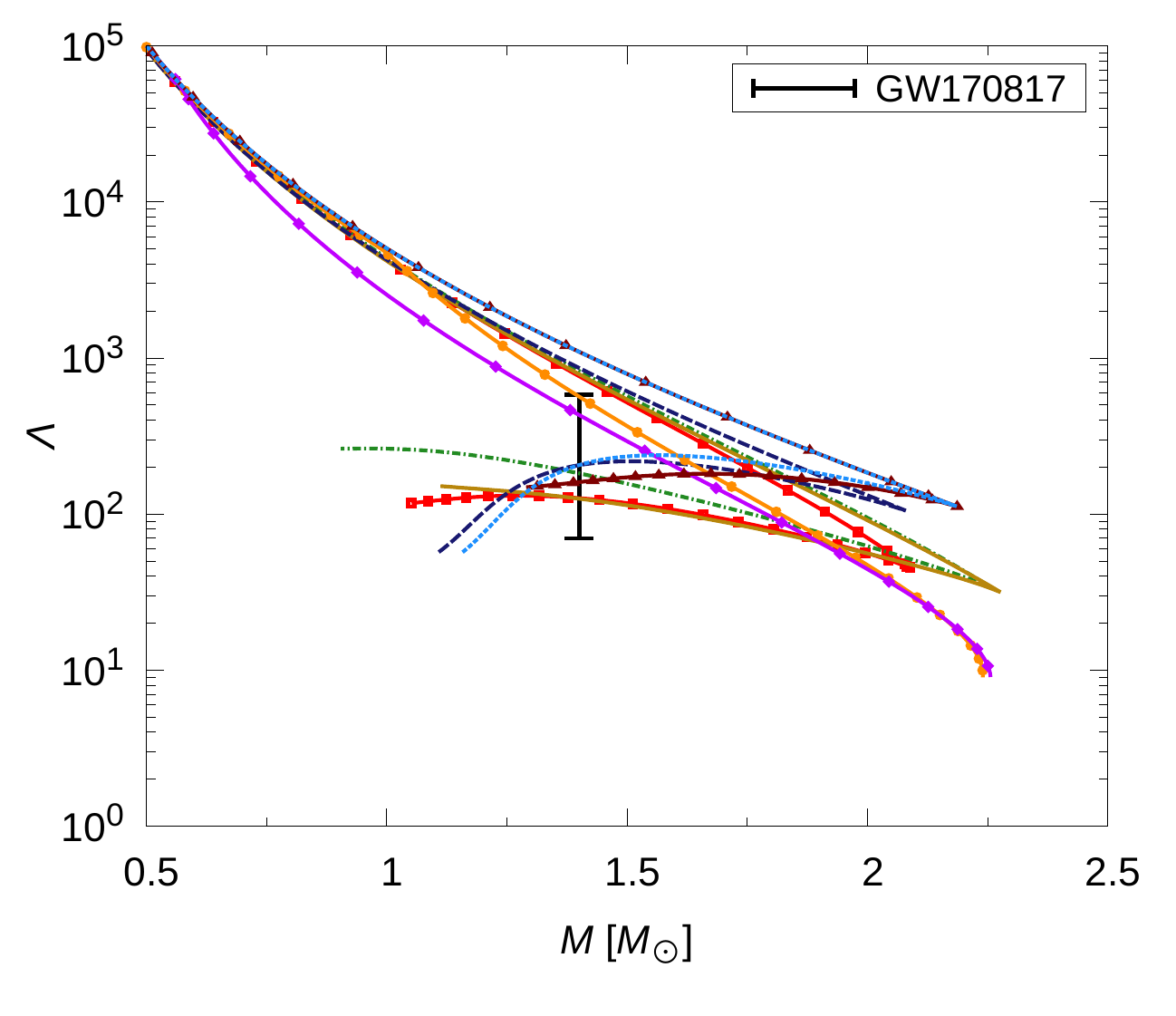}
\caption{Dimensionless tidal deformability $\Lambda$ as a function of the gravitational mass $M$ for the EOSs of Fig.~\ref{fig:equation-of-state} (same colors and symbols are used). In all cases, the hybrid branch fulfils the constraint from the GW170817 event \cite{Abbott:2018wiz} but the hadronic branch does not.
For slow conversions, all points of the curves represent stable stars and $\Lambda$ can be either an increasing or a decreasing function of $M$. For rapid conversions, the hybrid branches of models $1$-$4$, $7$ and $8$ are unstable.}
\label{fig:Mlambda}
\end{figure}

\begin{figure}
\centering
\includegraphics[width=0.6\linewidth,angle=0]{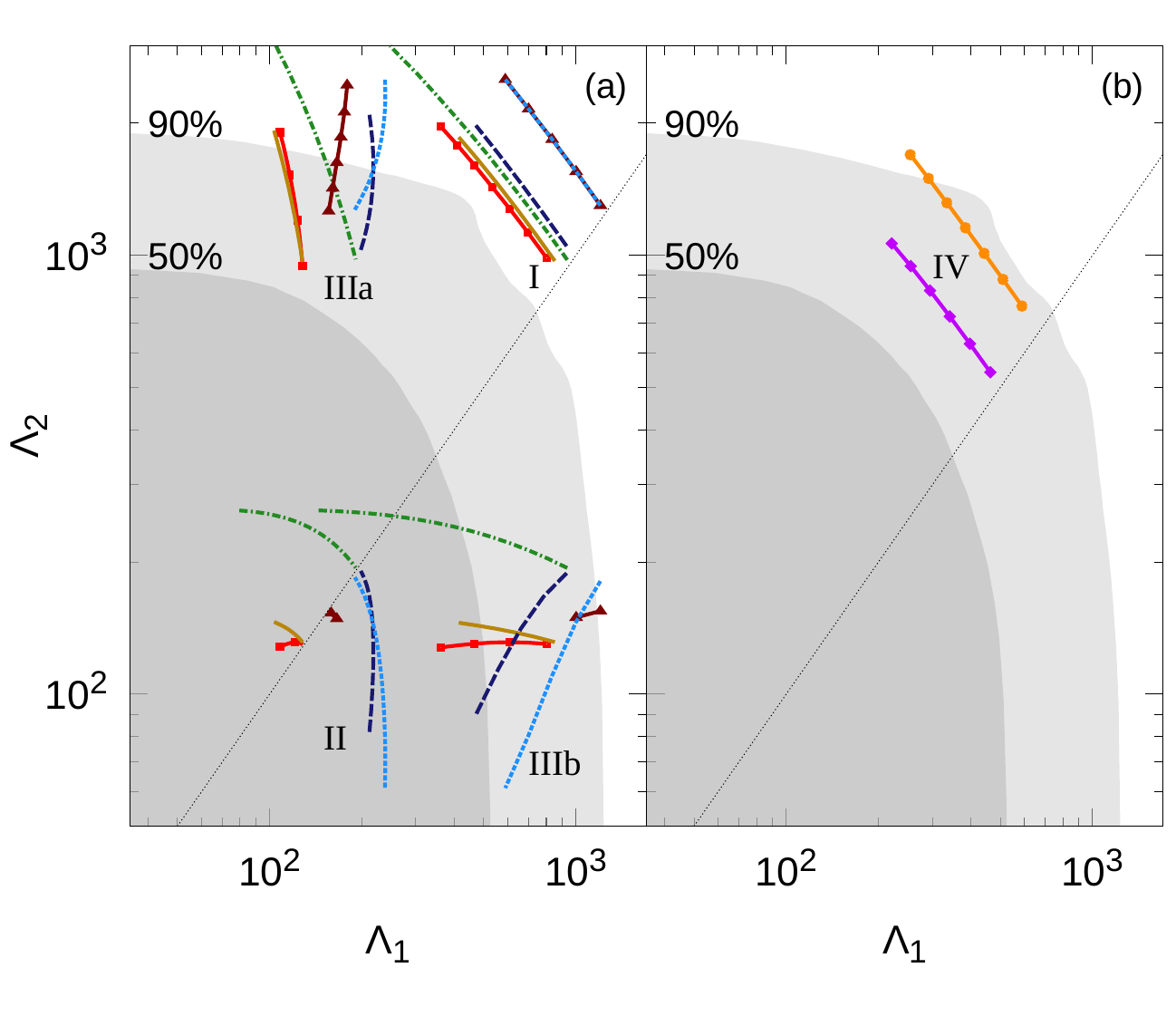}
\caption{Tidal deformabilities $\Lambda_1$-$\Lambda_2$ for BNSMs with the same chirp mass and mass ratio as GW170817 for the EOSs of Fig.~\ref{fig:equation-of-state} (same colors and symbols are used). Results are split in two panels according to the scenarios described in the text: (a) high-pressure and (b) low-pressure transition. 
Labels indicate the merging of two hadronic stars (I), two SSHSs (II), a hadronic star with a SSHS (IIIa, IIIb), and two totally stable HSs (IV).    
For slow conversions, $\Lambda_1 > \Lambda _2$ is possible because $\Lambda$ does not necessarily decrease with $M$.
}
\label{fig:Lambda1-Lambda2}
\end{figure}

In the following,  we present our results for the $M$-$R$ relationship (Fig.~\ref{fig:MR}) and the tidal deformability (Figs.~\ref{fig:Mlambda} and \ref{fig:Lambda1-Lambda2}) using the  EOSs of Fig. \ref{fig:equation-of-state}. In all figures, we show simultaneously the results for slow and rapid conversions. When slow conversions are assumed, all points of the thick curves in Fig.~\ref{fig:MR} and all points of the curves in Figs.~\ref{fig:Mlambda} and \ref{fig:Lambda1-Lambda2} represent stable configurations.  In the rapid case, all HSs with $\partial M / \partial \epsilon_c < 0$ are unstable (i.e. HSs of models $1$-$4$, $7$ and $8$).

Our results show that astrophysical constraints are satisfied in two different situations (see Fig.~\ref{fig:MR}): 

\emph{(a) Scenario with high-pressure transition} (models \mbox{$1$-$4$}, $7$ and $8$). A long SS hybrid branch is found to the left of the maximum mass point in the $M$-$R$ diagram and astrophysical constraints are fulfilled for a wide range of values of the quark EOS parameters. Twin stars with the same $M$ but different $R$ are possible for a wide range of masses. The density at the center of SSHSs can be as high as some tens of $n_0$ (e.g., model $2$ reaches $\sim 66\,n_0$ at the center of the last stable object).
In this scenario, the stiff hadronic branch fulfills NICER and $2~M_{\odot}$ pulsar constraints, and  SSHSs explain GW170817 as well as $2~M_{\odot}$ pulsars and PSR J0740+6620 observations. 
For a fixed hadronic EOS and a given $p_t$,  the length of the SS hybrid branch depends mostly on $\Delta \epsilon$ and on the stiffness of the quark EOS. Thus, a sufficiently large $\Delta \epsilon$ is needed at the interface for explaining GW170817.

\emph{(b) Scenario with low-pressure transition} (models $5$ and $6$).  The main feature of this case is the existence of a long hybrid stellar branch that is stable for both rapid and slow conversions. For this \emph{totally stable} hybrid branch the condition $\partial M / \partial \epsilon_c > 0$ is verified.
Observations are not easily fulfilled in this scenario when the \mbox{PSR J0740+6620} radius is taken into account and \emph{fine tuning} is needed, with low values of $p_t$ and $\Delta \epsilon$. 

The dimensionless tidal deformability $\Lambda$ is shown in Fig.~\ref{fig:Mlambda} as a function of $M$. When rapid conversions are assumed,  $\Lambda$ for stable configurations has the standard behavior \cite{Chatziioannou:2018vzf}, i.e. the larger the mass the smaller the $\Lambda$.  However, for slow conversions, $\Lambda$ can decrease or increase with $M$ meaning that, for a given hybrid EOS, not necessarily the most massive component of a binary NS merger (BNSM) will have the smallest $\Lambda$. As for the $M$-$R$ relationship, the GW170817 constraint is satisfied in the low and in the high pressure interface scenario.

In Fig.~\ref{fig:Lambda1-Lambda2}, we show the $\Lambda_1$-$\Lambda_2$ relationship spanning all possible combinations for the masses of the BNSM compatible with the GW170817 event. 
All BNSMs involving two hadronic stars are outside the 90\% confidence contour of GW170817 (label I in Fig.~\ref{fig:Lambda1-Lambda2}), which is an expected result due to our choice of extremely stiff hadronic EOSs. 
In the case of slow conversions many combinations are in agreement with GW170817: binaries with two SSHSs are inside the 50\% region (II), binaries involving a hadronic star and a SSHS are mostly inside the 90\% contour (IIIa and IIIb), and binaries with two totally stable HSs are inside the the 90\% region (IV).
If rapid conversions are assumed, only binaries involving two totally stable HSs fall inside the 90\% confidence contour (IV).

Finally, it is worth mentioning that calculations for \textit{fully-stable} hybrid stars were done in Refs. \cite{Paschalidis:2017qmb,Alvarez-Castillo:2017qki,Li:2021sxb,Li:2019fqe}. Notice that, although these authors do not indicate explicitly the conversion timescale at the interface, they use the derivative $\partial M / \partial \epsilon_c$ to asses stability. Consequently, their curves do not include the SSHSs studied in the present work.  The part of our curves corresponding to \textit{fully-stable} models is in agreement with the results of those works.

%%%%%%%%%%%%%%%%%%%%%%%%%%%%%%%%%%%%%%%%%%%%%%%%%%%%%%%%%%%%%%%%%%

%%%%%%%%%%%%%%%% FIGURE 4

%-------------------------------------------
\section{Discussion}
\label{sec:6}
%-------------------------------------------

In this work we explored new prospects for HS models emerging when the speed of quark-hadron interface conversions is taken into account, and confronted them with current observational constraints.

When rapid interface conversions are assumed, we find that stable HSs compatible with astrophysical observations are possible only if the discontinuity occurs at low enough pressures and the quark EOS parameters are fine-tuned. This occurs because we concentrated intentionally on HSs with very stiff hadronic EOSs and on quark EOSs with only one linear piece. The use of hadronic EOSs of intermediate stiffness and different  quark EOSs, may allow other fits to present observational data, albeit less easily than before the NICER measurement of PSR J0740+6620 \cite{Annala:2020efq}.

On the other hand, our results for HSs with slow interface conversions bring a novel view of the structure of compact stars, which is completely consistent with current astrophysical observations and nuclear/pQCD restrictions. In this new scenario, purely hadronic NSs made of extremely stiff nuclear matter would have large radii and masses exceeding that of observed $2~M_{\odot}$ pulsars. To its left in the $M$-$R$ diagram, there would be a long branch of HSs whose dynamic stability is possible, in spite of $\partial M/\partial \epsilon_c$ being negative, due to the slowness of quark-hadron interface reactions. Our goal in this work has not been to perform an exhaustive analysis of all possible hybrid models that agree with current constraints, but to show that this explanation is feasible and does not require a fine tuning of EOS parameters. Our results also draw attention to the great relevance of microphysical properties such as the surface tension, the curvature energy and reaction timescales, which can completely change our understanding of NS structure but cannot be encoded in $p(\epsilon)$ relationships and derived quantities, no matter how general or comprehensive they may be.

The probable existence of SSHSs opens new interesting scenarios in NS physics and astrophysics. For the hadronic matter EOS, it remarks that stiff and ultra stiff hadronic EOSs are still viable and compatible with current observations. For the quark matter EOS, the possibility of reaching densities tens of times greater than those normally expected in NSs, reinforces the astrophysical significance of studies that explore perturbative QCD in the low-temperature and high-density regime \cite{Kurkela:2009gj,Kurkela:2014vha,Gorda:2018gpy}. 
It also shows that there is not necessarily a tension between astrophysical observations and the theoretically expected conformal limit of the speed of sound \cite{Reed:2019ezm,Bedaque:2014sqa}:
large observed masses and radii would be explained by hadronic matter with very repulsive contributions and a large sound speed, while the small deformability of GW170817 is naturally explained by a high-pressure first-order phase transition to weakly interacting dense quark matter with $c_{s}^{2} \rightarrow 1 / 3$.
For astrophysics, a new scenario for the existence of two families of NSs is available, together with the standard hybrid star one \cite{Alford:2013aca,Benic:2014jia,Christian:2019qer,Shahrbaf:2019vtf} and the proposal of joint existence of hadronic and self-bound strange quark stars \cite{Bombaci:2004mt,Drago:2015cea}.  Compared to twin stars already studied in the literature, twins involving SSHSs may span a wider mass range, approximately between $\sim 1\, M_{\odot}$ to more than $2\,M_{\odot}$.

But how could the SSHS branch be populated in a realistic astrophysical scenario?  Many cold hadronic stars of models 1-4, 7 and 8 are in metastable states because it is energetically convenient for them to convert into more compact SSHSs with same baryonic mass.  However, these hadronic stars attain $p_t$ only at the center of the maximum mass object. Therefore, mass accretion onto them  would not be able to produce SSHSs because the object would collapse to a black hole when $p_t$ is attained at the stellar center.
However, in hot hadronic objects such as a protoneutron stars  \cite{Fischer:2017lag} or NSs created after a compact star merger \cite{Baiotti:2008ra,Weih:2019xvw} the conversion to quark matter can be triggered at the core at a pressure significantly smaller than $p_t$.  The reasons are the following.
On one hand, the QCD phase diagram suggests that the density of hadron-quark phase equilibrium gets smaller as the temperature, $T$, increases. Indeed, although the state-of-the-art understanding of the phase structure of QCD matter allows robust conclusions only at  finite temperature with a small density and at an asymptotically high density, there are several investigations of the whole phase diagram using effective models. Many of these models find that the phase transition from a deconfined quark phase to confined hadronic matter is of the first order at large chemical potential with a critical point at intermediate non-zero chemical potential at the end of the first order phase transition line. This generic behavior suggests that the hadron-quark transition gets easier as the temperature gets larger. 
On the other hand, if the conversion is triggered by quantum or thermal nucleation,  it has been shown that the critical stellar mass above which a metastable hadronic star could undergo a phase transition is significantly reduced when the object is hot. As shown in Figs. 14 and 15 of  Ref. \cite{Bombaci:2016xuj},  when the entropy per baryon of the protoneutron star is $\sim 2 k_B$, the critical stellar mass decreases by $\sim 10-20\%$ with respect to the cold star. Thus, a significant portion of the upper part of the hadronic branch of our models 1-4, 7 and 8 would be prone to a transition to the SSHS branch in the protoneutron star phase.   Hot post-merger hadronic stars would be even more propitious environments for conversion due to the high $T$ (up to $100$ MeV \cite{Bauswein:2018bma,Weih:2019xvw}) and the existence of large density fluctuations \cite{Baiotti:2008ra,Weih:2019xvw}.   
The above scenario, does not imply that purely hadronic objects with masses close to $M_{\mathrm{max}}$ cannot exist.  Below the ``hot'' critical mass, a proto-hadronic star would survive the early stages of its evolution without decaying to a SSHS. When the hadronic object cools down, the critical mass rises and the star can accrete additional mass from a companion keeping its hadronic nature. This scenario deserves further investigation, but it suggests that there are feasible channels for populating both branches.

To conclude, we mention some features of SSHSs that can help in their observational identification. Certainly, precise mass and radius measurements for a sufficiently large population of sources will significantly reduce the degeneracy of theoretical models and may open up the possibility of identifying SSHS branches if they exist and are long enough. Additionally, future GW detector networks will be able to measure the masses and tidal deformabilities to high accuracy, as well as some quasinormal mode frequencies to within tens of Hz  \cite{Pratten:2019sed}. The tidal deformability of SSHSs is significantly smaller and the $f$-mode frequency, $\nu_f$, considerably larger \cite{Tonetto:2020bie, mariani2022MNRAS} than the corresponding values of a hadronic star of the same mass. This characteristic is in agreement with claims that hyper-excited dynamical tides, i.e., anomalously small $\nu_f$, are disfavored by GW170817 \cite{Pratten:2019sed}. 
Moreover, discontinuity $g$-modes can be excited in SSHSs but don't exist in the case of rapid conversions due to the absence of a buoyancy force \cite{Tonetto:2020bie}. Contrary to $g$-modes of totally stable HSs which have $\nu_g \lesssim 1 \, \mathrm{kHz}$ and very long damping times,  $g$-modes of SSHSs have  $\nu_g \approx 1- 2 ~ \mathrm{kHz}$ and much shorter damping times that facilitate their detection for a given excitation amplitude \cite{Tonetto:2020bie,mariani2022MNRAS}. This property make SSHSs falsifiable by GW asteroseismology.

\section*{Acknowledgements}

G.L. acknowledges the support of the Brazilian agencies CNPq (grant 316844/2021-7) and FAPESP (grants 2022/02341-9 and 2013/10559-5). M.M. and I.F.R-S. thank CONICET and UNLP (Argentina) for financial support under grants PIP-0714 and G157, G007. IFR-S is also partially supported by PICT grant 2019-0366 from ANPCyT, Argentina and by the National Science Foundation (USA) under Grant PHY-2012152.

\bibliographystyle{JHEP}
\bibliography{letter}

\end{document}